\documentstyle[11pt,epsfig]{article}

\def\be{\begin{equation}}
\def\ee{\end{equation}}
\def\bea{\begin{eqnarray}}
\def\eea{\end{eqnarray}}
\def\ba{\begin{array}}
\def\ea{\end{array}}
\def\a{\alpha}

\def\d{\delta}

\def\0{$\Gamma_0$}

\def\p{\phi}

\def\t{\theta}

\def\L{${\cal L}$ }
\def\m{M\"obius }
\def\vt{\vartheta}

\title{Dimer statistics on the M\"obius strip and the Klein bottle}

\author{W. T. Lu and F. Y. Wu \\
Department of Physics\\
 Northeastern University\\
Boston,
Massachusetts 02115, USA}

\begin{document}

\maketitle

\medskip

\begin{abstract}
 Closed-form expressions are obtained for the generating function
of close-packed dimers on a $2M \times 2N$
simple quartic lattice embedded on a M\"obius strip and a Klein bottle.  
Finite-size corrections are also analyzed and
compared with those under cylindrical and free boundary conditions.
Particularly, it is found that, for large lattices of the same size and with a square
symmetry, the number of dimer configurations on a M\"obius strip 
is 70.2\% of that on a cylinder.  We also establish two identities
relating dimer generating functions for  M\"obius strips and   cylinders.
 
   \end{abstract}
\newpage

\section{Introduction}
 It is well-known that the problem of closed-packed dimers on  planar
graphs can be solved, meaning that its generating function
can be formulated as  a 
Pfaffian, the square root of an antisymmetric determinant \cite{kas}. 
This leads to the possibility that the generating function  can be evaluated
in a closed form for lattices with periodic structures. 
Indeed, Kasteleyn \cite{kas} has succeeded in obtaining the closed-form 
expressions of the
generating function for the simple quartic lattice
with both free and toroidal boundary conditions, and McCoy and T. T. Wu \cite{mccoywu}
extended the result to  cylindrical boundary conditions. 
 Temperley and Fisher \cite{tf}, and Fisher \cite{fisher} have also solved 
the free boundary case independently,
  although the generalization of the
field-theoretic approach used in \cite{fisher} to general graphs 
is less apparent.
  
\medskip
While the solutions  under
various boundary conditions yield the
same per-site generating function 
in the bulk limit, their finite-size corrections are nevertheless different.
  In this context there exists two unique 
surface structures, namely, the M\"obius strip and the Klein bottle,
 which have
escaped heretofore attention.  The \m strip
 is  a non-orientable surface          which has
  a {\it single} side and a {\it single}
 boundary edge, and the Klein bottle, also  non-orientable,  has
a single side and     {\it no} edges.
In view of the increasing interest on finite-size corrections, particularly
their interplay with the conformal field theory \cite{bcn}, it  is of pertinent interest to
study lattice models on these surfaces.
   In this {\it Letter} we consider this problem.
We solve exactly the dimer statistics for  a 
simple quartic net
embedded on a \m strip and on a Klein bottle.
We also analyze its finite-size corrections
 and establish two identities relating generating functions
 under the M\"obius and cylindrical boundary conditions.

\section{Formulation}
Consider first a $2M\times 2N$ simple quartic lattice ${\cal L}$ embedded on a
\m strip.  
Here, for convenience, we assume the number of
sites be even in each direction.
The \m strip of length $2N$ and width $2M$ is shown in Fig. 1,
 where  repeated sites are denoted by open circles.
 We consider  close-packed dimers on ${\cal L}$.

\begin{figure}[htbp]
\center{\rule{5cm}{0.mm}}
\rule{5cm}{0.mm}
\vskip -1.4cm
\hskip 4.5cm
\epsfig{figure=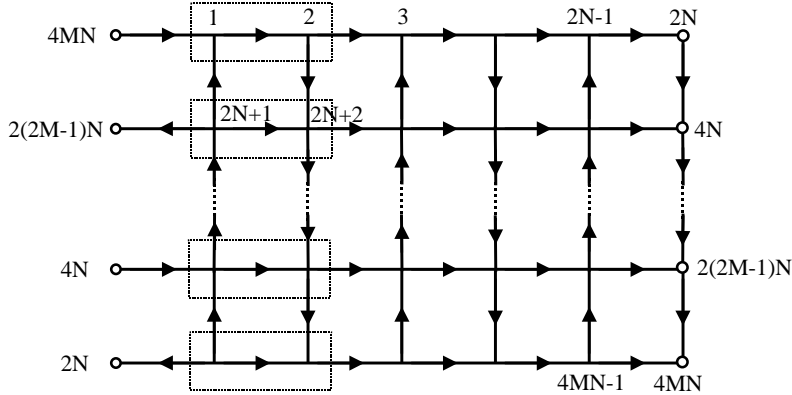,height=6.0in}
\vskip -3.7in
\caption{A M\"obius strip of a simple quartic net of length $2N$ and width $2M$.
Sites denoted by open circles are repeated as indicated by the labeling.
The edge orientation 
satisfies the Kasteleyn  clockwise-odd rule for superposition polygons,
and rectangles of dotted lines denote dimer cities.
\label{fig:fig1}}
\end{figure}

\medskip
Let the dimer weights be $z_1$ and $z_2$, respectively, for dimers in the
horizontal and vertical directions.  We need to   evaluate
the dimer generating function
\be
Z^{Mob}_{{(2M, 2N)}}  (z_1,z_2)
 = \sum_{\rm dimer\>config.}z_1^{n_h}z_2^{n_v}, \label{gen}
\ee
where $n_h$ and $n_v$ are, respectively, the numbers of horizontal and vertical
dimers subject to the sum rule
$n_h +n_v  = 2MN$.
The summation in (\ref{gen}) is taken over all close-packed dimer configurations.

\medskip
A key element in the Pfaffian formulation of the generating function 
 is to orient edges of \L such that every superposition polygon,
formed by superimposing  two dimer configurations,
must contain an odd number of arrows in the clockwise direction,
the {\it clockwise-odd} rule.  Kasteleyn \cite{kas} has shown that
when this rule is satisfied, then the generating function of close-packed
dimers can be written as a Pfaffian.  For the \m strip ${\cal L}$, 
it can be shown  that
the clockwise-odd rule is satisfied
if we adopt the edge orientations shown in Fig. 1.  Then,  
we have the result
 \be
Z^{Mob}_{(2M,2N)}(z_1,z_2)= {\rm Pf}(A) = \sqrt {{\rm det}| A|},
\ee
where $A$ is a $4MN\times 4MN$ antisymmetric matrix whose elements
can be read off from Fig. 1.

 To see the general structure of $A$,   it is instructive to first
write down 
the matrix $A$ in a simpler case.
For this purpose  consider    the $2\times 4$
\m strip ($M=1, N=2$)   shown in Fig. 2.  Reading off from Fig. 2, we obtain
\be
A=\pmatrix{0&z_1&0&0&-z_2&0&0&-z_1\cr
-z_1&0&z_1&0&0&z_2&0&0\cr
0&-z_1&0&z_1&0&0&-z_2&0\cr
0&0&-z_1&0&-z_1&0&0&z_2\cr
z_2&0&0&z_1&0&z_1&0&0\cr
0&-z_2&0&0&-z_1&0&z_1&0\cr
0&0&z_2&0&0&-z_1&0&z_1\cr
z_1&0&0&-z_2&0&0&-z_1&0\cr}
\ee
and hence
\be
Z^{Mob}_{(2,4)} (z_1,z_2)= 2\>z_1^4 + 4 \>z_1^2z_2^2 + z_2^4, \label{gen24}
\ee
a result which can be verified by explicit enumerations.

\begin{figure}[htbp]
\center{\rule{5cm}{0.mm}}
\rule{5cm}{0.mm}
\vskip -1.3cm
\hskip 4.5cm
\epsfig{figure=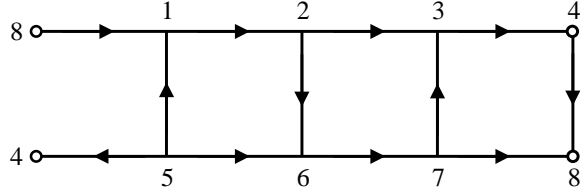,height=6.0in}
\vskip -4.8in
\caption{A $2\times 4$ \m strip.
\label{fig:fig2}}
\end{figure}

\medskip
To write down $A$ for general $M$ and $N$,
 we adopt ``dimer cities"
formed by pairing sites $\{2n, 2n+1\}, n=1,2,\cdots,2MN$
as indicated in Fig. 1, and introduce 
$2\times 2$ matrices
\bea
a(0,0) &=& \pmatrix{0 & z_1 \cr -z_1 & 0}, \hskip 1cm
a(0,1) = \pmatrix{-z_2 & 0 \cr 0& z_2} \nonumber \\
a(1,0) &=& \pmatrix{0 & 0 \cr z_1 & 0}, \hskip 1.5cm
a(-1,0) = \pmatrix{0 &-z_1 \cr 0 & 0}. \label{amatrices}
\eea
Further introduce $N\times N$ matrices
\bea
F_N &=& \pmatrix {0 & 1 & 0 &\cdots & 0 \cr
                  0 & 0  & 1 &\cdots & 0 \cr
                  \vdots & \vdots & \vdots & \ddots & \vdots \cr
                  0 & 0  & 0 &\cdots & 1 \cr
0 & 0  & 0 &\cdots & 0},
K_N = \pmatrix {0 & 0 & 0 &\cdots & 0 \cr
                  0 & 0  & 0 &\cdots & 0 \cr
                  \vdots & \vdots & \vdots & \ddots & \vdots \cr
                  0 & 0  & 0 &\cdots & 0 \cr
1 & 0  & 0 &\cdots & 0},
\eea
and the $2M\times 2M$ matrix
 \be
J_{2M} = \pmatrix { &    &  & & & & 1 \cr
                          & 0 &  & & & -1 &  \cr
                   &     &  & & 1 &  &  \cr
                   &     & & -1 & &  &  \cr
                   &   &   \cdots & & &  &  \cr
                    & 1    & & & &  0&  \cr
                 -1 &  &   & & &  &  }.
\ee
Then, the matrix $A$  can be written in terms
of a linear combination of direct products of smaller ones,
 \be
A = A_{2N}\otimes I_{2M} +a(0,1) \otimes I_N \otimes \biggr[F_{2M} -F_{2M}^T\biggr]
+B_{2N} \otimes J_{2M},
\label{mobiusA}
\ee
where
\bea
A_{2N}& =&a(0,0) \otimes I_N + a( 1,0) \otimes F_N
+a(-1,0)\otimes F_N^T \nonumber \\
B_{2N} &=&-a(1,0) \otimes K_N + a(-1,0) \otimes K^T_N,\label{ab}
\eea
 $F^T_N$ and $K^T_N$  are transposes of $F_N$ and $K_N$, 
and $I_N$ is the $N\times N$ identity matrix.
 Note that the first two terms in (\ref{mobiusA}) are 
the corresponding $A$ matrix 
for  free boundary conditions. 

We next consider a Klein bottle constructed
by connecting the upper and lower edges
of the \m strip of Fig. 1 
  in a periodic fashion with $2N$ 
extra vertical edges.  The clockwise-odd rule
 again holds if these edges are oriented inversely to those in the same
column.
 Thus, we have the matrix
\be
A^{Klein}= A -a(0,1)
\otimes I_{N} \otimes (K_{2M}-K_{2M}^T).\label{kleinmatrix}
\ee
  
\section{Evaluation of  determinants}
We first evaluate det$|A|$. 
Since $J_{2M}$ commutes with $F_{2M} -F_{2M}^T$, and as a result
 they can be
diagonalized by applying
 a common similarity transformation.
Introducing  the $2M\times 2M$ matrix $U$  whose elements are
\bea
U_{m,m'} &=& i^m\sqrt{2\over {2M+1}}\> \>\sin 
\biggl( {{mm'\pi}\over {2M+1}}\biggr) \nonumber \\                
(U^{-1})_{m,m'}& =&  (-i)^{m'}\sqrt{2\over {2M+1}} \>\sin
 \biggl( {{mm'\pi}\over {2M+1}}\biggr) ,\nonumber \\
&& \hskip 3cm m,m'= 1,2,...,2M,
\eea
we find
 \bea
(U^{-1}I_{2M} U)_{m,m'}& = & \d_{m,m'} \nonumber \\
(U^{-1}J_{2M} U)_{m,m'}&=& i\ (-1)^{M+m} \ \d_{m,m'} \nonumber \\
 (U(F_{2M} -F_{2M}^T)U^{-1})_{m,m'} &=& (2i\cos \phi_m)\ \d_{m,m'},
\hskip 0.5cm
m=1,2,...,2M, \label{eigen}
\eea                 
 where 
\be
\phi_m = {m\pi\over( {2M+1})}.
\ee
Thus, we can replace the $2M\times 2M$ matrices in (\ref{mobiusA}) 
 by their
respective eigenvalues,    and express det$|A|$ as a product of
the replaced determinants,
namely,
\be
{\rm det}|A|=\prod_{m=1}^{2M} {\rm det} \biggl| A_{2N} + \bigl(2i\cos \phi_m\bigr) a(0,1)
\otimes I_N +(-1)^{M+m} i B_{2N} \biggr|. \label{ab1}
\ee
 Introducing (\ref{ab}) and the $N\times N$ matrix
\be
T_N= F_N+i(-1)^{M+m+1}K_N = \pmatrix{  0 & 1 & 0 &\cdots & 0 \cr
                  0 & 0  & 1 &\cdots & 0 \cr
                  \vdots & \vdots & \vdots & \ddots & \vdots \cr
                  0 & 0  & 0 &\cdots & 1 \cr
i(-1)^{M+m+1} & 0  & 0 &\cdots & 0},
\ee
we can rewrite (\ref{ab1}) as
\be
{\rm det}|A|=\prod_{m=1}^{2M} {\rm det} \biggl| \bigl[a(0,0) +
\bigl(2i\cos \phi_m\bigr) a(0,1) \bigr] \otimes I_N + a(1,0) \otimes T_N
+ a(-1,0)\otimes T_N^\dagger \biggr|,
  \label{ab2}
\ee
where $T_N^\dagger$ is the Hermitian conjugate of $T_N$.

Now $T_N$ and $T_N^\dagger$ commute and 
they can be simultaneously diagonalized
with respective eigenvalues $e^{i\t_n}$ and 
$e^{-i\t_n}$, where
\be
\t_n = (-1)^{M+m+1}
\biggl[{{(4n-1)\pi} \over {2N}}\biggr], \hskip 1cm n=1,2,...,N.
\ee
It follows that we have
\bea
Z^{Mob}_{(2M,2N)}(z_1,z_2)&=&\sqrt{{\rm det}|A|}\nonumber \\
&=&\prod_{m=1}^{2M} \prod_{n=1}^N\biggl[{\rm det} \biggl| a(0,0) +
\bigl(2i\cos \phi_m \bigr) a(0,1) \nonumber \\
&& \hskip 3cm  + a(1,0) e^{i\t_n}
+ a(-1,0) e^{-i\t_n}   \biggr| \biggr]^{1/2}\nonumber \\
&=& \prod_{m=1}^M \prod_{n=1}^N\biggl[ 4 z_1^2 \sin^2{{(4n-1)\pi}\over 
{4N}} +
 4 z_2^2 \cos^2{{m\pi}\over {2M+1}} \biggr],
  \label{ab3}
\eea
where we have used  $\cos^2 \phi_m =  \cos^2 \phi_{2M -m+1}$
to effect  the square root. 
It is readily verified that (\ref{ab3}) reduces to
(\ref{gen24})  when $M=1, N=2$.

The 
determinant det$|A^{Klein}|$ of the matrix (\ref{kleinmatrix}) can be evaluated
in a similar fashion    from which  we obtain
\be
 Z^{Klein}_{(2M,2N)}(z_1,z_2)=\prod_{m=1}^M\prod_{n=1}^N
 \Bigg[4z_1^2\sin^2{(4n-1)\pi\over 4N}+4z_2^2\sin^2{{(2m-1)\pi}\over {2M}}
 \Bigg].  \label{klein}
\ee
As a comparison,  the generating functions
 for $2M\times 2N$  lattices with 
free boundaries \cite{kas} and cylindrical boundary conditions (periodic
in the $N$ direction) \cite{mccoywu}
are, respectively,
\bea
 Z^{free}_{(2M,2N)}(z_1,z_2)&=&\prod_{m=1}^M\prod_{n=1}^N
 \Bigg[4z_1^2\cos^2{n\pi\over 2N+1}+4z_2^2\cos^2{m\pi\over 2M+1}
 \Bigg],\nonumber \\
 Z^{cyl}_{(2M,2N)}(z_1,z_2)&=&\prod_{m=1}^M\prod_{n=1}^N
 \Bigg[4z_1^2\sin^2{(2n-1)\pi\over 2N}+4z_2^2\cos^2{m\pi\over 2M+1}
 \Bigg].\label{freecyl}
 \eea
The similarity between  expressions ({\ref{ab3}) -  ({\ref{freecyl})
is striking,
with the 
 only difference  being the trigonometric factors which can be
associated to the respective boundary conditions.
We remark that for  $(2M-1)\times 2N$ lattices we have \cite{mccoywu}
\be
Z^{cyl}_{(2M-1,2N)}(z_1,z_2)={1\over{2z_1^N}} \prod_{m=1}^M\prod_{n=1}^N
 \Bigg[4z_1^2\sin^2{(2n-1)\pi\over 2N}+4z_2^2\cos^2{m\pi\over 2M}
 \Bigg].\label{freecyl1}
 \ee

These expressions give rise to the same bulk ``free energy" 
\bea
f_{\rm bulk} &=& \lim_{M,N\to\infty}(4MN)^{-1} \ln Z_{(2M,2N)}(z_1,z_2) \nonumber \\
  &=&{1\over {16\pi^2}} \int_0^{2\pi} d\t\int_0^{2\pi} d\phi \ln (4z_1^2
\sin^2\t +
  4 z_2^2 \cos^2\phi).\label{bulk}
\eea
Particularly, the number of ways $W_{8\times 8}$ that an $8 \times 8$ 
checkerboard can be covered
by 32 dominoes is obtained by setting $M=N=4$ and $z_1=z_2=1$, yielding
\bea
W_{(8,8)}^{free} &=& \ 12\ 988\ 816 = 2^4\times (901)^2 \nonumber \\
W_{(8, 8)}^{Mob} &=& \ 46\ 029\ 729 = 47 \times 271 \times 3617 \nonumber \\
W_{(8, 8)}^{cyl} &=& \ 71\  385\ 601 = (8449)^2 \nonumber \\
W_{(8,8)}^{Klein} &=&    220\ 581\ 904 = 2^4\times (3713)^2 \nonumber \\
W_{(8,8)}^{toroidal} &=& 311\ 853\ 312 = 2^8\times  3^2\times 135353,
\eea
where  the number for toroidal boundary conditions is computed 
using the result of \cite{kas}.

{\it Two identities}: There exist curious identities connecting
generating functions for  M\"obius strips and cylinders.
Brankov and Priezzhev \cite{bp} have proposed the identity
 \be
Z^{cyl}_{(2M,4N)}(z_1,z_2) = \biggl[ Z^{Mob}_{(2M,2N)}(z_1,z_2)\biggr]^2, 
\label{identity}
\ee
which  they established up to the order of $c_3$
in the large $M,N$ expansion (\ref{finitesize}). 
The identity (\ref{identity}) can now be 
rigorously established using  (\ref{ab3}) and
(\ref{freecyl}).  The proof 
 follows from the
fact that in the product over $n=1,2,\cdots,4N$ on the LHS, 
the factor $\sin ^2[(2n -1)\pi/4N]$ reproduces the set 
$\sin ^2[(4n' -1)\pi/4N]$, $n'=1,2,\cdots, N$, exactly twice. 
 For $(2M-1) \times 2N$ lattices we have instead
the  identity
  (see {\it Note added} below) 
\be
Z^{cyl}_{(2M-1,4N)}(z_1,z_2) = {1\over  2}\biggl[ Z^{Mob}_{(2M-1,2N)}(z_1,z_2)\biggr]^2. \label{identity1}
\ee

\section{Finite-size corrections}
For {\it large}  $M$ and $N$,
we expect to have
\be
\ln Z_{(2M,2N)} = 4MN  f_{\rm bulk} + 2Nc_1+ 2Mc_2 
+c_3 + O(1/N), \label{finitesize}
\ee
where 
$c_1$, $c_2$ and $c_3$ are constants dependent at most on 
\be 
\tau = z_2/z_1 \hskip 1cm {\rm and} \hskip 1cm \xi=N/M,
\ee
the aspect ratio.
The computation of finite-size corrections for products of the form
of (\ref{ab3}) is standard \cite{fisher,barber,f},
and is outlined in the following.

We first take out a factor $z_1^{2MN}$ in (\ref{ab3})
  and carry out the product over $n$  by introducing the identity \cite{gr}
 \be
\prod_{n=1}^{N}\Bigg[x^2-2\cos\Bigg({2n\pi\over N}-\a\Bigg)+x^{-2}\Bigg]=x^{2N}-2\cos N\a+x^{-2N},
\ee
with $\alpha = \pi/2N$.
This leads  to 
\be
Z_{(2M,2N)}^{Mob}(z_1,z_2) = z_1^{2MN} \prod_{m=1}^M \biggl[ x^{2N}(\tau,\phi_m)
+ x^{-2N}(\tau,\phi_m)\biggr],  \label{xx}
\ee
where
\be
x^{2}(\tau,\phi)  +x^{-2}(\tau,\phi) =2 + 4 \tau^2 \cos^2 \phi,
\ee
 or, explicitly,
\be
x(\tau,\phi) = \tau   \cos \phi + \sqrt{ 1+ \tau^2 \cos^2 \phi}.
\ee
It follows that we have
\be
\ln Z^{Mob}_{(2M,2N)} = 2MN \ln z_1 + 2N  \sum_{m=1}^M \ln x(\tau,\phi_m)
  + \sum_{m=1}^M \ln \biggl[ 1 + x^{-4N}(\tau,\phi_m) \biggr].  \label{fsum}
\ee

We next carry out the first summation in (\ref{fsum}) using the Euler-MacLaurin summation
formula 
\bea
\sum_{m=1}^Mf(a+m\d)&=&{1\over \d}\int_{a}^{a+M\d}f(\p)d\p
+{1\over 2}\biggl[f(a+M\d)-f(a)\biggr]\nonumber \\
&&\quad+{\d\over 12}\biggl[f'(a+M\d)-f'(a)\biggr]+O(\d^3),
\eea
with $a=0$ and $\d=\pi/(2M+1)$.
The first  term and the leading contribution of the second term lead to
 the following expressions for the  bulk free energy,
\bea
f_{\rm bulk}
 &=& {1\over 2} \ln z_1 + {1\over \pi}\int_0^{\pi/2} 
\ln x(\tau,\phi) d \phi \nonumber  \\
 &=& {1\over 2} \ln z_2 + {1\over \pi}\int_0^{\pi/2} 
\ln x(\tau^{-1},\phi) d \phi ,
\label{free}
\eea
where the second line is written down by symmetry.

After some manipulation following \cite{barber}, 
this leads to
the following   for the M\"obius boundary condition:
\bea
c_1^{Mob}(\tau) &=&{1\over \pi} \int_0^{\pi/2} \ln \biggl[{{x(\tau,\phi)}\over
 {x(\tau,0)}}\biggr]
d \phi \ <0
\nonumber \\
c_2^{Mob}(\tau) &=& 0 \nonumber \\
c_3^{Mob}(\tau,\xi)& =&\ln\vt_3\bigl({\pi/ 4},q\bigr)
+ {1\over 24} \pi \tau \xi -\sum_{\ell=1}^\infty \ln (1-q^{2\ell}),\label{theta}
\eea
where $\vt_3$ is the theta function
\be
\vt_3(u,q)= 1+2 \sum_{n=1}^\infty q^{n^2} \cos (2nu),
\ee
with $q=e^{-\pi \xi\tau/2}$. Note that 
$c_2^{Mob}=0$  is a consequence of the fact that there is no
boundary edge in the $M$ direction.

For the cylindrical boundary condition periodic in the $N$ direction,
  one has in place of (\ref{xx}),
\be
Z_{(2M,2N)}^{cyl}(z_1,z_2) = z_1^{2MN} \prod_{m=1}^M \biggl[ x^{2N}(\tau,\phi_m)+2
+ x^{-2N}(\tau,\phi_m)\biggr],
\ee
from which one obtains
\bea
c_1^{cyl}(\tau)&= &c_1^{Mob}(\tau), \hskip 1cm c_2^{cyl}(\tau)=0 \nonumber \\
c_3^{cyl}(\tau,\xi)&= &c_3^{Mob}(\tau,\xi) + \ln \bigl[ {{\vt_3(0,q)} \big/
{ \vt_3\bigl({\pi/ 4},q\bigr)} }\bigr].
\eea
It follows that, for large $M,N$, we have
\be
{{Z_{(2M,2N)}^{Mob}(z_1,z_2)}\over {Z_{(2M,2N)}^{cyl}(z_1,z_2)}}
= {{\vt_3\bigl({\pi\over 4},q\bigr)}\over {\vt_3(0,q)} }.
\label{ratio}
\ee
Particularly, for $M=N$ and $z_1=z_2 =1$,  we have $q=e^{-\pi/2}$ and 
 (\ref{ratio}) gives the ratio of the numbers of dimer configurations 
under the two respective boundary conditions
as 0.701845. 
  
For free boundary conditions we obtain
\bea
c_1^{free}(\tau) &=&  c_1^{Mob} (\tau), \hskip 1cm
c_2^{free}(\tau) =  c_1^{Mob} (\tau^{-1}) \nonumber \\
 c_3^{free}(\tau,\xi ) &=&  
c_3^{Mob}(\tau, \xi)+c_1^{Mob}(\tau)
+{1\over 2} \ln
\biggl[ {{2\sqrt{1+\tau^2}}\over {1+\sqrt{1+\tau^2}}}\biggr]  . \label{c2}
\eea
Here, use has been made  of  the second line of (\ref{free}).
  These expressions  reproduce those obtained
by Fisher \cite{fisher} and Ferdinand \cite{f}.
Similarly we have for the Klein bottle 
\bea
c_1^{Klein} (\tau)&=& c_2^{Klein} (\tau)=0 \nonumber \\
c_3^{Klein} (\tau, \xi)&=& c_3^{Mob} (\tau,\xi)+ \pi \tau \xi/8.
\eea

If one takes the limit of $N\to \infty$ ($M\to \infty$) first while keeping
$M$ ($N$) finite \cite{bcn}, one obtains from (\ref{finitesize}) 
\bea
\lim_{N\to\infty} {1\over {2N}} Z_{(2M,2N)}
&=& 2M f_{bulk} + c_1 +{ {\Delta_1}/ {2M}}
+O(1/M^2),\label{Nlimit} \nonumber \\
\lim_{M\to\infty} {1\over {2M}} Z_{(2M,2N)}
&=& 2N f_{bulk}  +  c_2 +{ \Delta_2/ {2N}}
+O(1/N^2). \label{Mlimit}
\eea
One finds after some algebra the results

\begin{center} 
\begin{tabular}{cccccc} \\
\hline\hline
 & M\"obius &Klein & Free & Cylindrical & Toroidal \\
\hline
$c_1$ & $c_1^{Mob}(\tau)$& 0& $c_1^{Mob}(\tau)$& $c_1^{Mob}(\tau)$&0 \\
$c_2$ &0  &0 &$c_1^{Mob}(1/\tau)$& 0 &0 \\
$\Delta_1$ & $ \pi \tau /24$ & $ \pi \tau/6 $& 
$\pi \tau /24$ & $\pi \tau /24$ & $ \pi \tau/6 $ \\
$\Delta_2 $&  $\pi /24\tau $ & $ \pi  
/24\tau$  & $\pi /24 \tau$ & $\pi /6\tau$ & $ \pi/6 \tau$\\
\hline\hline
\end{tabular}
\end{center}

\noindent
 where we have also included results for toroidal boundary conditions
 worked out from results of \cite{f}.
The tabulated values of $\Delta_1$ and $\Delta_2$ 
are consistent with the finding 
 of \cite{bcn} that the   periodical boundary condition
(in the lateral direction of an infinite strip) is distinct
from  the free or fixed,  now including the M\"obius, boundary conditions. 
While it is tempting to extract the central charge $c$ from 
$\Delta_1$ and $\Delta_2$ resulting in
$c=\tau$
and $\tau^{-1}$, respectively, 
 we remark that  the extraction is superfluous since 
results of \cite{bcn} apply to systems
at criticality while the present dimer system does not  
exhibit a critical point.

{\it Note added}:
After the submission of this paper
we learned of a recent preprint 
by Tesler \cite{gt} in which the dimer generating function $Z^{Mob}_{({\cal M},{\cal N})}
(z_1,z_2)$ 
is enumerated
for   all ${{\cal M},{\cal N}}$ (even or odd) as linear combinations
of two Pfaffians
with the results given
in terms of a $q$-analogue of the Fibonacci number.
We have verified that for ${{\cal M},{\cal N}}$ even Tesler's result 
is the same as our expression (\ref{ab3}).
 For  ${\cal M}$ odd and ${\cal N}$ even, Tesler found the solution  also
 given by (\ref{ab3}) but with an extra factor  $z_1^{-{\cal N}/2}$ with 
 $2M\to {\cal M}$, $2N\to {\cal N}$, and
 the product over $m$  taken  from  $1$ to $({\cal M}+1)/2$. We remark that 
 this
  leads to the identity (\ref{identity1}) upon using (\ref{freecyl1}).
 For  ${\cal M}$ even and ${\cal N}$ 
odd, however, the solution assumes a more complicated
expression \cite{gt}.
 The M\"obius strip problem has also been studied
by Brankov and  Priezzhev \cite{bp} in the 
context of the large $M,N$  expansion (\ref{finitesize}) 
of the free energy.

  \section{Acknowledgement}
  We are grateful to R. P.
Stanley for calling our attention to Ref. \cite{gt} and
 V. B. Priezzhev   for pointing out to us  Ref. \cite{bp}.
 We are also indebted to  P. Kleban for
an enlightening discussion on the connection with the conformal field theory.
This work has been supported in part by NSF Grant
DMR-9614170.



\begin{thebibliography} {9}
\bibitem{kas} 
P. W. Kasteleyn,  Physica, 27 (1961) 1209.
\bibitem{mccoywu}
For a 
comprehensive review on dimer statistics using Pfaffians see
B. M. McCoy and T. T. Wu,   
{\it Two-dimensional Ising model} (Harvard
University Press, Cambridge, Massachusetts, 1973)

 

\bibitem{tf} H. N. V. Temperley  and M. E. Fisher, Phil. Mag. 6 (1961) 1061. 
 \bibitem{fisher} 
M. E. Fisher, Phys. Rev. 124 (1961) 1664.
 \bibitem{bcn} H. W. J. Bl\"ote, J. L. Cardy  and M. P. Nightingale,
 Phys. Rev. Lett. 56 (1986) 742.
\bibitem{barber} M. N. Barber and B. W. Ninham, {\it Random and restricted
walks}, (Gordon and Breach, New York 1970), Appendix AI.
\bibitem{f} A. D. Ferdinand, J. Math. Phys. 8 (1967) 2332.
\bibitem{gr} I. S. Gradshteyn and I. M. Ryzhik, {\it Table of integrals,
series, and products}, (Academic Press, New York 1994) 1.394.
\bibitem{gt} G. Tesler, {\it Matchings in graphs on non-oriented 
surfaces}, preprint. 
\bibitem{bp} J. G. Brankov and V. B. Priezzhev, Nucl. Phys.  B400(FS) (1993) 633.

 \end{thebibliography}
\end{document}